\input{aipcheck}

\edef\optionlist{%
   \variorefoptionifavailable        
   draft,%
   \psnfssproblemoption              
   tnotealph}

\documentclass[\optionlist]{aipproc}

\newcommand{\be}{\begin{equation}}
\newcommand{\ee}{\end{equation}}
\newcommand{\bea}{\begin{eqnarray}}
\newcommand{\eea}{\end{eqnarray}}

\def\simg{{\ \lower-1.2pt\vbox{\hbox{\rlap{$>$}\lower6pt\vbox{\hbox{$\sim$}}}}\ }}
\def\siml{{\ \lower-1.2pt\vbox{\hbox{\rlap{$<$}\lower6pt\vbox{\hbox{$\sim$}}}}\ }}

\layoutstyle{6x9}

\usepackage{ttct0001}

\usepackage{shortvrb}
\MakeShortVerb\|

\makeatother

\begin{document}

\begin{flushright}
{\small CERN-TH/2001-242\\
August 2001}
\end{flushright}

\author{Tobias Hurth}{
  address={CERN, Theory Division, 
        CH 1211 Geneva 23, 
        Switzerland},
  email={tobias.hurth@cern.ch},
}

\author{Thomas Mannel}{
    address={
Institut f\"ur Theoretische Teilchenphysik, 
Universit\"at  Karlsruhe,  D--76128 Karlsruhe, Germany}
}

\title{Direct CP Violation In Radiative $B$ Decays}

\date{}

\begin{abstract}
    We discuss the role of the radiative $B$ decays $B \rightarrow 
    X_{s/d} \, \gamma$  in our search for new physics focusing on 
    the issue of direct CP violation.
    We discuss in some detail a SM prediction for the CP asymmetries 
    in inclusive 
    $b \rightarrow s/d$ transitions, namely 
    $\mid \Delta {\cal B}(B \to X_s \gamma) + 
    \Delta {\cal B}(B \to X_d \gamma) 
   \mid \sim 1 \cdot 10^{-9}$. Any measured value in significant deviation 
    of this estimate 
    would indicate new sources of $CP$ violation beyond the CKM phase.\\
 
\begin{center}
{{\it  Invited talks given by T.H. at the IPPP Workshop 
'Phenomenology of Physics Beyond the SM`, Durham (England), 
6-11 May 2001, and at the  International Workshop on QCD 'QCD@Work`, 
Martina Franca (Italy), 16-20 June 2001; to appear in the Proceedings.}}
\end{center}

\end{abstract}

\maketitle

\section{Inroduction}

It is well-known that radiative 
$B$ decays like $B \rightarrow X_{s/d} \,  \gamma$ 
play an important role in our search of new physics. 
These processes test the SM directly on the quantum level and, thus, 
are particularly sensitive to new physics (for a recent review see 
\cite{Review}). 
While the direct production of new (supersymmetric) 
particles is reserved for the planned 
hadronic machines such as the LHC at CERN, the indirect search of the 
$B$ factories  already implies significant restrictions for the parameter 
space of supersymmetric models and will thus lead to important 
clues for the direct search of supersymmetric particles.
It is even possible that these rare processes lead to the first 
evidence of new physics by a significant deviation from the SM
prediction, for example in the observables concerning direct 
CP  violation, although it will then be difficult to identify
in this way the new structures in detail. But also in the long run,
after new physics has already been discovered, these decays will
play an important role in analyzing in greater detail the 
underlying new dynamics.

One of the main difficulties in examining the observables in 
$B$  physics is the influence of the strong interaction.
As is well known, for matrix elements dominated by long-distance strong 
interactions there is no
adequate quantitative solution available in quantum field theory.
The resulting hadronic uncertainties restrict the opportunities in
$B$ physics significantly. 
If new physics does not show up in $B$ physics 
through large deviations as recent experimental data indicates
the focus on theoretically clean variables like inclusive
radiative $B$ decays is mandatory.

Within inclusive decays like  $B \rightarrow X_{s/d} \,  \gamma$ 
the long-distance strong interactions are
less important and well under control due to the heavy mass expansion.
The decay  $B \rightarrow X_s \gamma$ was first observed by 
the CLEO collaboration
\cite{Cleoinclusive}; these measurements have been 
refined \cite{Cleoinclusive2,Cleoinclusive3} and confirmed by 
other experiments  \cite{Alephinclusive,Belleinclusive} (see also 
 \cite{Romeinclusive}).
The present world average  
using the present data from BELLE, CLEO and ALEPH is
\begin{equation} 
\label{worldaverage}
{\cal B}(B \to X_s \gamma) = (3.22 \pm 0.40) \times 10^{-4}.
\end{equation}

The theoretical prediction of the Standard Model (SM)
up to  next-to-leading logarithmic (NLL) precision
for the total decay rate of the $B \rightarrow X_s \, \gamma$ mode 
\cite{NLL}
is well in agreement with the experimental data.
The theoretical NLL prediction  for the
$B \rightarrow  X_s \gamma$ branching ratio \cite{greubhurthQCD}
leads to   
\begin{equation}
{\cal B}(B \rightarrow X_s \gamma) = (3.32 \pm 0.14 \pm 0.26)\times 10^{-4},
\label{currentprediction}
\end{equation}
where the first error represents the uncertainty 
regarding the scale dependences, while the second error is the
uncertainty due to the input parameters. In the second error 
the uncertainty due to the parameter $m_c/m_b$ is dominant.

This inclusive mode already 
allows for theoretically clean and rather strong constraints on 
the parameter space of various extensions of the 
SM \cite{NLLbeyond,NLLbeyond2,NLLbeyond3}.
While many phenomenolgical analyses of the inclusive 
$B \rightarrow X_{s/d} \gamma $ decays in supersymmetry are retricted to 
the assumption of minimal flavour
violation including only CKM induced flavour change,
a model-independent analysis also has to consider the generic new  
supersymmetric sources of flavour change due to the mixing in the squark mass 
matrix. In \cite{NLLbeyond3} new model-independent bounds on 
    supersymmetric flavour-violating parameters were derived from 
     $B \rightarrow X_s \gamma$. 
The importance of 
interference effects 
for the bounds on the 
parameters in the squark mass matrices within the unconstrained MSSM
is explicitly demonstrated.  
In former analyses no 
correlations between the different sources of flavour violation 
were taken into account \cite{NLLbeyond2}.
The new bounds are in general one order of magnitude weaker 
than the original bounds 
on the single off-diagonal element, which 
was derived in previous work \cite{NLLbeyond2} by
neglecting any kind of interference effects.

\begin{figure}
\includegraphics[height=0.30\textwidth,angle=0]{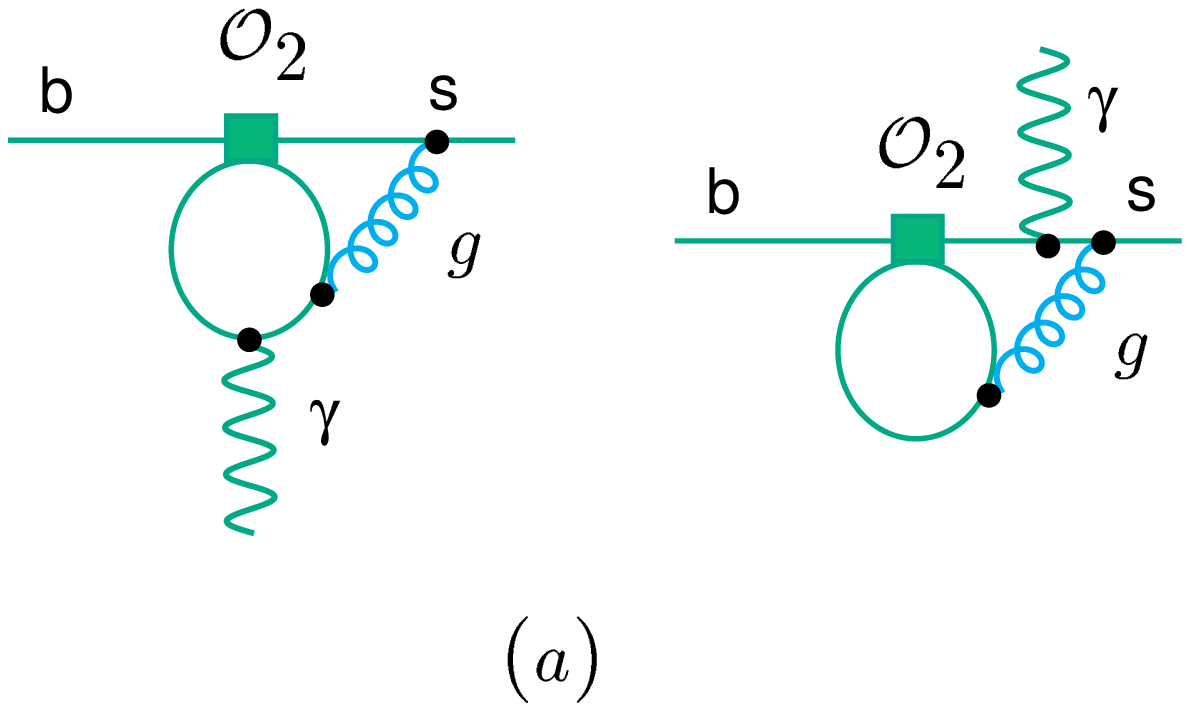}
\hspace{2cm}\includegraphics[height=0.30\textwidth,angle=0]{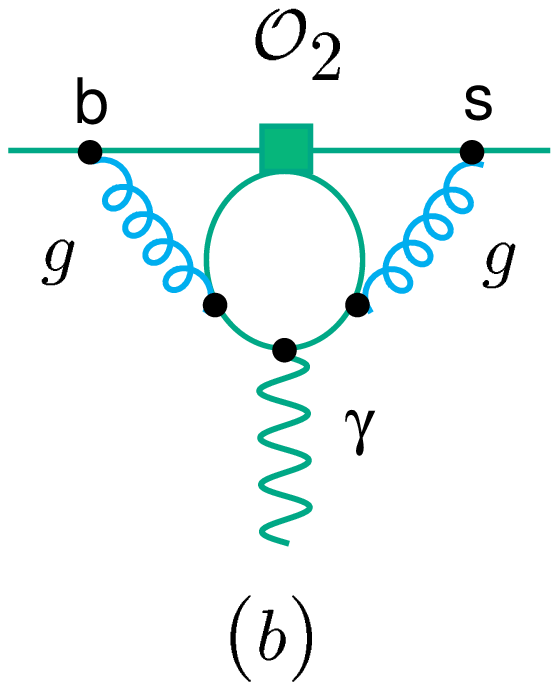}
\caption{
{\it a)
Typical diagrams contributing 
to the matrix element of the operator ${\cal O}_2$ at the NLL level.
b) Typical diagram contributing to the NLL anomalous 
dimension matrix.
}}
\label{Feynman}
\end{figure}

Recently, quark mass effects within the decay 
$B \rightarrow X_s \gamma$ were further analysed \cite{GambinoMisiak}, 
in particular the definitions of the quark masses $m_c$ and $m_b$. 
The charm quark enters in specific NLL matrix elements 
(see Fig. \ref{Feynman}) where the charm quark
mass is dominantly off-shell. Therefore, 
the authors of \cite{GambinoMisiak} argue 
that the running charm mass should be chosen instead
of the pole mass.                
The latter choice was used  
in all previous analyses \cite{Towards,NLL,CDGG,Kagan,greubhurthQCD}.  
\begin{equation}
 m_c^{\rm pole}/m_b^{\rm pole} \qquad 
{{\Rightarrow}}\, \qquad   
m_c^{\overline{\rm MS}}(\mu)/m_b^{\rm pole}, \,  \,\,
\mu \in [m_c,m_b].
\end{equation}
Since these  matrix elements start at NLL order only and, thus, 
the renormalization scheme for $m_c$ and $m_b$ is an NNLL issue,
one should regard this choice as an educated guess of the NNLL
corrections. 
The new choice is guided by the 
experience gained from many higher-order calculations
in perturbation theory.  
Numerically, the shift from $m_c^{\rm pole}/m_b^{\rm pole} =  0.29 \pm 0.02$
to $m_c^{\overline{\rm MS}}(\mu)/m_b^{\rm pole} = 0.22 \pm 0.04$
is rather important and leads to a $+ 11 \%$ shift of the central 
value of the $B \rightarrow X_s \gamma$ branching ratio.  
With their new choice of the charm mass 
their theoretical prediction 
for the branching ratio is \cite{GambinoMisiak}
\begin{equation} 
\label{totalbr}
{\cal B}(B \to X_s \gamma) = (3.73 \pm 0.30) \times 10^{-4},
\end{equation}
which induces a small difference between the theoretical and the experimental 
value. 

However, because the choice of the renormalization scheme for $m_c$ and $m_b$ 
is a NNLL effect, their important 
observation should be reinterpreted into a
larger error bar in $m_c^{\overline{\rm MS}}(\mu)/m_b^{\rm pole}$ 
which includes also the value of $m_c^{pole}$.
A  more conservative choice would then be  
$m_c^{\overline{\rm MS}}(\mu)/m_b^{\rm pole} = 0.22 \pm 0.07$ which 
deletes  the significance of the perceived discrepancy.

\section{Direct CP Asymmetry}

The CP violation  in the $B$ system 
will yield an important independent test of the SM description of 
CP violation (see \cite{Durham}). In particular, detailed measurements
of CP asymmetries in rare $B$ decays will be possible in the near 
future. Theoretical predictions for the {\it normalized} CP asymmetries of
the  inclusive channels (see~\cite{AG7,KaganNeubert,SoniWu}) within 
the SM lead to 
\begin{equation}
\label{SMpredict} 
\nonumber 
\alpha_{CP}({B \rightarrow  X_{s/d} \, \gamma}) =
\frac{\Gamma(\bar B \rightarrow X_{s/d}\gamma)
     -\Gamma(B \rightarrow  X_{\bar s/\bar d}\gamma)}
     {\Gamma(\bar B \rightarrow  X_{s/d} \gamma)
     +\Gamma(B \rightarrow  X_{\bar s/\bar d}\gamma)}
\end{equation}
\begin{equation}                    
  \alpha_{CP}({B \rightarrow  X_s \gamma}) \approx 0.6 \%, \qquad 
  \alpha_{CP}({B \rightarrow B_d \gamma}) \approx  -16 \%
\label{SMnumbers}
\end{equation}
when the best-fit values for the CKM parameters \cite{CKMfit} 
are used.
An analysis for the leptonic counterparts  is presented in
\cite{Hiller}.
The normalized CP asymmetries may also be calculated for
exclusive decays; however, these results are model-dependent. An
example of such a calculation may be found in \cite{GSW}.

CLEO has already presented a measurement of the CP asymmetry in
inclusive $b \to s \gamma$ decays, yielding \cite{CleoCP}
\begin{equation}
\alpha_{CP}(B \rightarrow  X_s \gamma) = 
(-0.079 \pm 0.108 \pm 0.022) \times (1.0 \pm 0.030) \, , 
\end{equation}
which indicates that very large effects are already excluded.

Supersymmetric predictions for the CP asymmetries in 
$B \rightarrow X_{s/d} \gamma$ depend strongly on what is
assumed for the supersymmetry-breaking sector and are, thus,  
a rather model-dependent issue. 
The minimal supergravity model cannot account for large 
CP asymmetries beyond $2\%$ because of the constraints coming 
from the electron and neutron electric dipole moments
\cite{Goto}. However, more general models 
allow for larger asymmetries,  of the order of $10 \%$ 
or even larger \cite{Aoki,KaganNeubert}.
Recent studies of the $B \to X_d \gamma$ rate asymmetry  
in specific models led to asymmetries between $-40 \% $ and $+40\%$
\cite{Recksiegel} or  $-45 \%$ and $+ 21 \%$ \cite{Asatrian}.
In general, CP asymmetries may lead to clean 
evidence for  new physics by a significant deviation from the SM
prediction.

In \cite{mannelhurth} it was explicitly derived, that
a bound on the {\it combined} asymmetries within  
the decays $b \to s \gamma$ and
$b \to d \gamma$, as well as their leptonic counterparts is possible
which is more suitable for the experimental settings.  
It provides a
stringent test, if the CKM matrix is indeed the only source of
CP violation. 
Using  U-spin, which is the $SU(2)$ subgroup of flavour $SU(3)$ relating
the $s$ and the $d$ quark and which is already a well-known tool 
in the context of nonleptonic decays \cite{Fleischer:1999pa,GronauBCP4},
one derives  relations between the CP asymmetries of the exclusive 
channels $B^- \to K^{*-} \gamma$ and $B^- \to \rho^- \gamma$ and 
of the inclusive channels $B \rightarrow X_s \gamma$ and 
$B \rightarrow X_d \gamma$.
One should make use  of the U-spin symmetry only
with respect to the strong interactions.
Moreover, within {\it exclusive} final states,
the vector mesons like the U-spin doublet ($K^{*-}$, $\rho^-$) 
are favoured as final states because 
these have masses much larger than the (current-quark) masses
of any of the light quarks. Thus one expects, for the ground-state 
vector mesons, the U-spin symmetry to be quite accurate in spite of the
nondegeneracy of $m_d$ and $m_s$. 
Defining the rate asymmetries ({not} the {\it normalized} CP asymmetries)
by  
\begin{equation} \label{ratediff}
\Delta \Gamma (B^- \to V^- \gamma) = \Gamma (B^- \to V^- \gamma) - \Gamma (B^+ \to V^+ \gamma)
\end{equation}
one arrives at the following relation \cite{mannelhurth}:
\begin{equation} 
\label{SMSM}
 \Delta \Gamma (B^- \to K^{*-} \gamma) +
\Delta \Gamma (B^- \to \rho^- \gamma) = b_{exc} \Delta_{exc}
\end{equation}
where the right-hand side is written as a product of a relative 
U-spin breaking $b_{exc}$ and a typical size  $\Delta_{exc}$ 
of the CP violating rate difference. 
In order to give an estimate of the right-hand side, one can use
the model result  from \cite{GSW}  for $\Delta_{exc}$,
\begin{equation}
\Delta_{exc}  = 2.5 \times 10^{-7}\,\,  \Gamma_B.
\end{equation}
The relative breaking $b_{exc}$ of U-spin can be estimated, 
e.g. from spectroscopy. This leads us to 
\begin{equation}
\mid b_{exc} \mid  =  \frac{M_{K^*} - m_\rho}{\frac{1}{2}(M_{K^*} + m_\rho)}
= 14 \%. 
\end{equation}
Certainly, other estimates are also possible, such as a comparison of
$f_\rho$ and $f_{K^*}$. In this case one finds 
a very small U-spin breaking. Using the 
more conservative value for
$b_{exc}$, which is also compatible with sum rule calculations of form factors 
(see  \cite{Ali:1994vd}), one arrives at
the standard-model prediction for the
difference of branching ratios
\begin{equation} \label{resexc1}
\mid \Delta {\cal B}(B^- \to K^{*-} \gamma) +
\Delta {\cal B}(B^- \to \rho^- \gamma) \mid \sim 4 \times 10^{-8}
\end{equation}
Note that the right-hand side is model-dependent.

Quite recently, the U-spin breaking effects were also estimated 
in the QCD factorization approach \cite{buchalla}.
Within this approach, it was shown that the U-spin breaking effect
essentially scales with the differences of the two form factors,
$(F_{K^*}-F_\rho)$.
Using the formfactors from the QCD sum rule calculation in 
\cite{ballbraun} and maximizing the CP asymmetries by a specific
choice of the CKM angle $\gamma$, the authors of \cite{buchalla} obtain
\begin{equation} 
 \Delta {\cal B}(B^- \to K^{*-} \gamma) +
\Delta {\cal B}(B^- \to \rho^- \gamma)  \sim - 3 \times 10^{-7},
\end{equation}
while for the separate asymmetries they obtain:
\be
\Delta {\cal B}(B\to K^*\gamma) = - 7 \times 10^{-7}, \quad \quad
\Delta {\cal B}(B\to \rho\gamma) =  4 \times  10^{-7}. 
\ee
This calculation explicitly shows the limitations of the relation
(\ref{SMSM}) as a SM test. 

The issue is much more attractive in the {\it inclusive} modes.
Due to the  $1/m_b$ expansion for the inclusive process,
 the leading contribution is
the free $b$-quark decay. In particular, there is no sensitivity to
the spectator quark and thus one arrives at the following relation
for the CP rate asymmetries
\cite{mannelhurth}:
\begin{equation} \label{resincg1}
\Delta \Gamma (B \to X_s \gamma) +
\Delta \Gamma (B \to X_d \gamma) = b_{inc} \Delta_{inc}. 
\end{equation}

In this framework one relies on
parton-hadron duality; 
so one can actually compute the breaking 
of U-spin by keeping a nonvanishing strange quark mass. 
The typical size of $b_{inc}$ can be roughly 
estimated to be of the order of 
$\mid b_{inc} \mid \sim m_s^2/m_b^2 \sim 5 \times 10^{-4}$;  
$\mid \Delta_{inc}\mid$  is again the 
average of the moduli of the two CP rate
asymmetries. These have been calculated (for vanishing strange quark mass),
e.g. in  \cite{AG7} and thus one arrives at 
\begin{equation} \label{resinc3}
\mid \Delta {\cal B}(B \to X_s \gamma) +
\Delta {\cal B}(B \to X_d \gamma) \mid \sim 1 \times 10^{-9}. 
\end{equation}
Any measured value in significant deviation of (\ref{resinc3}) 
would be an indication of new sources of CP violation. Although 
only an estimate is given here, it should again  be stressed  
that in the inclusive mode the
right-hand side in (\ref{resinc3})  can be computed in a
model-independent way with the help  of the heavy mass expansion.

Going beyond leading order in the $1/m_b$ expansion the first subleading
corrections are of 
order $1 / m_b^2$ only.
The 
$1/m_b^2$ corrections are induced by the imaginary part 
of the forward scattering amplitude $T(q)$:
\be
\label{forward}
T(q) = i \, \int d^4x \, < B \mid T {\cal O}_7^+(x) \, {\cal O}_7(0) \mid
 B > \, \exp (iqx) \quad .
\ee
where only the magnetic operator ${\cal O}_7$ is taken into account. 
Using the operator product expansion for $T {\cal O}_7^+(x) \, {\cal O}_7(0)$
and heavy quark effective theory methods, the decay width
$\Gamma(B \to X_s \gamma)$ reads \cite{Falk,Alineu}
(modulo higher terms in the
$1/m_b$ expansion):
\bea
\label{width}
\Gamma_{B \to X_s \gamma}^{({\cal O}_7,{\cal O}_7)} &=&
\frac{\alpha  G_F^2 m_b^5}{32 \pi^4} \,  \mid V_{tb} V_{ts} \mid^2 \, C_7^2(m_b) \,
\left( 1 + \frac{\delta^{NP}}{m_b^2} \right) \quad , \nonumber \\
\delta^{NP} &=& \frac{1}{2} \lambda_1 - \frac{9}{2} \lambda_2 \quad ,
\eea
where $\lambda_1$ and $\lambda_2$ are the parameters 
for kinetic energy and the chromomagnetic
energy. Using $\lambda_1=-0.5 \, \mbox{GeV}^2$ and 
$\lambda_2=0.12 \, 
\mbox{GeV}^2$, one gets $\delta^{NP} \simeq -4\%$.
Thus, the contributions are small and
cancel in the sum of the rate asymmetries - in the limit of U-spin symmetry. 
The U-spin breaking effects in this  contribution  also 
induce an overall factor $m_s^2/m_b^2$ in addition - as one can read off 
from the explicit results for the $B \rightarrow X_s l^+l^-$ case
\cite{hillerali}.

\begin{figure}
\includegraphics[height=0.25\textwidth,angle=0]{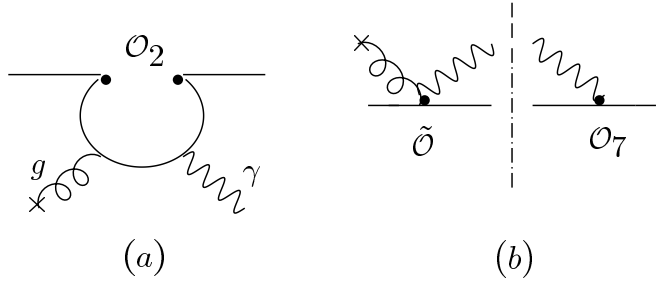}
\caption{
{\it a) Feynman diagram from which the operator $\tilde{\cal O}$
arises. b) Relevant cut-diagram for the 
$({\cal O}_2,{\cal O}_7)$-interference.}}
\label{Voloshinfig}
\end{figure}

There are also nonperturbative corrections
which scale with  $1/m_c^2$ \cite{voloshinwyler}. 
which are induced by the interference of the mangnetic ${\cal O}_7$ and 
the four-quark operator ${\cal O}_2$.
This effect is generated by the diagram in Fig. \ref{Voloshinfig}a 
(and by the one, not shown, where the gluon and the photon
are interchanged); 
$g$ is a soft gluon interacting with the
charm quarks in the loop. Up to a characteristic Lorentz structure,
this loop is given by the integral
\be
\label{volloop} 
\int_0^1 dx \, \int_0^{1-x} dy \, \frac{xy}{m_c^2-k_g^2 x(1-x) -2xy k_g 
k_\gamma}
\quad .
\ee
As the gluon is soft, i.e. $k_g^2,k_g k_\gamma  \approx \Lambda^{QCD} \, m_b/2
\ll m_c^2$, the integral can be expanded in $k_g$. The (formally)
leading operator, denoted by $\tilde{\cal O}$, is
\be
\label{tildeo}
\tilde{\cal O} = \frac{G_F}{\sqrt{2}} V_{cb} V_{cs}^* C_2 \,
\frac{e Q_c}{48 \pi^2 m_c^2} \, \bar{s} \gamma_\mu (1-\gamma_5) g_s 
G_{\nu \lambda} b \, \epsilon^{\mu \nu \rho \sigma} \partial^\lambda
F_{\rho \sigma} \quad .
\ee 
Then working out the cut diagram shown in Fig. \ref{Voloshinfig}b,
one obtains the nonperturbative 
contribution $\Gamma^{(\tilde{\cal O},{\cal O}_7)}_{B \to X_s \gamma}$
to the decay width,
which is due to the $({\cal O}_2,{\cal O}_7)$ interference.
Normalizing this contribution by the LL partonic width, one obtains
\be
\label{voleffect}
\frac{\Gamma^{(\tilde{\cal O},{\cal O}_7)}_{B \to X_s \gamma
}}{\Gamma_{b \to s \gamma}^{LL}} = -\frac{1}{9} \, \frac{C_2}{C_7} 
\frac{\lambda_2}{m_c^2} \simeq +0.03 \quad .
\ee
This result corresponds to the leading term in an expansion in the parameter
\mbox{$t=k_g k_\gamma / 2 m_c^2$}.
The expansion parameter is approximately 
$m_b \Lambda_{QCD}/2 m_c^2 \approx 0.3$ (rather than $\Lambda^2_{QCD}/m_c^2$)
and  it is not a priori clear
whether formally higher order terms in the $m_c$ expansion are
numerically suppressed. 
However, the explicit expansion of the complete vertex function, 
corresponding to  Fig. \ref{Voloshinfig}a, 
shows that  higher order terms are indeed suppressed, because
the corresponding coefficients are small (see i.e. \cite{rey}).

Moreover, the operator $\tilde{\cal O}$ does not  
contain any information on 
the strange mass, thus, also in these contributions one finds the same
 overall  suppression factor  from the U-spin breaking.

The corresponding long-distance contributions 
from up-quark loops are CKM suppressed
in the $B \rightarrow X_s \gamma$ case, but  this does not
hold in the    $B \rightarrow X_d \gamma$ case.
Naively, one could expect that the corresponding contributions
from up-quark loops scale with $1/m_u^2$. However, 
following  the approach of \cite{rey}, one easily shows that 
the general vertex function   
cannot be expanded in the parameter $t$ in that case. However, 
the expansion in inverse powers of $t$ is reasonable. 
The leading term in this expansion scales like $t^{-1} \sim m_u^2/k_g k_\gamma$ and therefore cancels the factor $1/m_u^2$ in the prefactor
(see the analogous  $1/m_c^2$ factor in (\ref{tildeo}))
and one gets  a  suppression factor $(\Lambda_{QCD}^2/m_u^2) \cdot 
(m_u^2/{k_g k_\gamma})$ \cite{rey}. Thus, although 
the  expansion in inverse
powers in $t$ induces  nonlocal operators, one explicitly finds 
that the leading term scales with $\Lambda_{QCD} / m_b$~\footnote{Also in the 
exclusive case   
an explicit  analysis within the 
QCD factorization approach leads to this suppression factor \cite{buchalla}.}.

This argument improves the discussion in \cite{mannelhurth} where the 
argument was given that vector dominance calculations in \cite{LDUP} show
that the long-distance contributions from the up-quark loops 
to the decay rates  are found to be rather small.
However, that argument in \cite{mannelhurth} does not allow  any 
statement about the effects  of  the U-spin breaking in constrast
to the one given here. 

Summing up, the analysis shows that the known nonperturbative contributions
to (\ref{resinc3}) are under control and small and that 
this prediction 
provides  a clean  SM test, if generic new CP phases are present or not.
Any significant deviation from the estimate (\ref{resinc3}) would be a 
strong hint to non-CKM contributions to CP violation.

Finally, we emphasize that an analogous prediction for the leptonic 
inclusive $B \rightarrow X_{s/d} \gamma$ modes is also possible taking into
account some specific cuts on the invariant dilepton spectrum.

\section*{Acknowledgements}
We are  grateful to  Gerhard Buchalla for discussions and  to Gino Isidori
for a careful reading of the manuscript.

\end{document}